\documentclass[sn-mathphys,Numbered]{sn-jnl}

\usepackage{graphicx}%
\usepackage{multirow}%
\usepackage{amsmath,amssymb,amsfonts}%
\usepackage{amsthm}%
\usepackage{mathrsfs}%
\usepackage[title]{appendix}%
\usepackage{xcolor}%
\usepackage{textcomp}%
\usepackage{manyfoot}%
\usepackage{booktabs}%
\usepackage{algorithm}%
\usepackage{algorithmicx}%
\usepackage{algpseudocode}%
\usepackage{listings}%
\theoremstyle{thmstyleone}%
\theoremstyle{thmstyletwo}%

\theoremstyle{thmstylethree}%

\begin{document}

\title[Article Title]{On the mass assembly history of the Milky Way: clues from its stellar halo}

\author*[1,2]{\fnm{Danny} \sur{Horta}}\email{dhortadarrington@flatironinstitute.org}

\author[1]{\fnm{Ricardo P.} \sur{Schiavon}}\email{R.P.Schiavon@ljmu.ac.uk}

\affil[1]{\orgdiv{Liverpool John Moores University}, \orgname{Organization}}

\affil[2]{\orgdiv{Centre for Computational Astrophysics}, \orgname{Flatiron Institute}}

\abstract{Stellar halos of galaxies retain crucial clues to their mass assembly history. It is in these galactic components that the remains of cannibalised galactic building blocks are deposited. For the case of the Milky Way, the opportunity to analyse the stellar halo's structure on a star-by-star basis in a multi-faceted approach provides a basis from which to infer its past and assembly history in unrivalled detail. Moreover, the insights that can be gained about the formation of the Galaxy not only help constrain the evolution of our Milky Way, but may also help place constraints on the formation of other disc galaxies in the Universe. This paper includes a summary of work undertaken during a PhD thesis aiming to make progress toward answering the most fundamental question in the field of Galactic archaeology: \textit{``How did the Milky Way form?''} Through the effort to answer this question, we summarise new insights into aspects of the history of assembly and evolution of our Galaxy and measurements of the structure of various of its Galactic components.}

\keywords{Galactic archaeology; Galaxy formation and evolution; Stellar haloes; Globular clusters}

\maketitle

\section{The Galactic stellar halo: a mosaic of the Milky Way's building blocks}\label{sec2}

Galaxies large and small are all predicted to grow and evolve under the same $\Lambda$ cold dark matter ($\Lambda$CDM) paradigm: hierarchical mass assembly. During this process of merging and growth, the baryonic material comprising the systems that get stripped and dissolved are typically deposited in the stellar halo of the resulting larger mass host. Therefore, stellar haloes of galaxies are mosaics comprised by a plethora of systems accreted over a galaxy's lifetime (i.e., the building blocks), whose stars encode the clues for rewinding the processes of assembly that led to the formation of galaxies we observe today. Moreover, in addition to providing clues to the formation and evolution of galaxies, studies of stellar halo populations can also help place constraints in many other realms of astrophysics (e.g., globular cluster formation and evolution, near-field cosmology, genesis of the elements, dark matter, amongst others).

Although small in mass fraction when compared to the total stellar mass of the Milky Way ($\lesssim1\%$), the Galactic stellar halo is a treasure-trove of relics that recently, thanks to the advent of data from stellar surveys in an unprecedented scale, has begun to reveal itself, and in turn the mass assembly history of the Galaxy. So far, studies of the Galactic stellar halo have shown that the bulk of the Milky Way halo formed early ($\gtrsim8$ Gyr), implying that the Galaxy's major building blocks are many gigayears old. It has been shown that the outer regions of the Milky Way's stellar halo are dominated by either more recent and/or lower mass debris from accreted satellite galaxies (e.g., \textsl{Sgr dSph}: \cite{Ibata1994}; \textsl{Cetus}: \cite[][]{Newberg2009}; \textsl{Orphan-Chenab} \cite{Grillmair2006,Belokurov2007} amongst others) or the infall of massive satellites (namely, the \textsl{LMC/SMC}), whereas the the inner regions host the debris from the major building blocks that constitute the bulk of the mass of the present day stellar halo and have shaped the formation of the Galaxy.

One of the most exciting results from \textsl{Gaia} was the unambiguous detection of a stellar halo substructure conjectured to be the remnant of an ancient and relatively massive accreted galaxy (the \textsl{Gaia-Enceladus/Sausage}: \citealp[][]{Belokurov2018,Helmi2018}\footnote{We note that there has been a plethora of work in the literature that have hinted at the detection of debris from a major accretion event \cite[e.g.,][]{Brook2003,Nissen2010,Hayes2018,Haywood2018,Mackereth2019}.}). This population has been shown to dominate the inner ($6\lesssim r_{\mathrm{GC}} \lesssim25$ kpc) regions of the stellar halo (\citealp[e.g.,][]{Deason2019,Iorio2019,Mackereth2020}), and is believed to be the culprit for the formation of the present day \textit{in situ} halo via heating of the old Milky Way disc (\citealp[e.g.,][]{Bonaca2017,DiMatteo2019,Belokurov2020}). In addition, several other discoveries of debris from accreted systems have been advanced (see \cite{Helmi2020} for a review), including the \textsl{Heracles} structure that we will discuss in the next section \citep{Horta2021}. While the nature and reality of these stellar halo populations requires further examination to be fully established, all these findings have helped elucidate the chronicle of the Milky Way. 

 In this paper, we summarise a body of work from a PhD thesis aimed at tackling a fundamental question in Galactic astronomy: \textit{How did the Milky Way form?} Armed with: 1) multifaceted big Milky Way data supplied primarily from the synergy of the revolutionary \textsl{Gaia} mission and \textsl{APOGEE} survey, and 2) state-of-the-art cosmological simulations of Milky Way-mass galaxies (\textsl{FIRE}, \textsl{EAGLE}), in the following we present a summary of results focused on unravelling the accretion history of the Milky Way (Section~\ref{sec_acchist}); grounding observational findings on the Milky Way with expectations from theory using cosmological simulations (Section~\ref{sec_cossims}); and the role of star clusters in the mass assembly history of the Galaxy and the formation of its stellar halo (Section~\ref{sec_gcs}).

 \section{The accretion history of the Milky Way} \label{sec_acchist}

The precision astrometry delivered by \textsl{Gaia} \citep[]{Gaia2018,Gaiaedr3}  and detailed chemical compositions supplied by massive spectroscopic surveys (e.g., \textsl{APOGEE} \cite{Majewski2017}) have led to a revolution in our understanding of the structure and properties of the Galactic stellar halo. This in turn has led to a deeper understanding of the accretion history of the Milky Way. Equipped with these data, several groups have proposed the discovery of distinct halo substructures conjectured to be the debris from different accreted systems during the Milky Way's lifetime \cite{Ibata1994,Belokurov2018,Helmi2018,Koppelman_thamnos,Myeong2019,Naidu2020}. While the identification of these halo substructures has helped constrain our understanding of the mass assembly history of the Milky Way, their association with any particular accretion event still needs to be clarified. This is largely because predictions from numerical simulations suggest that a single accretion event can lead to multiple substructures in phase space \cite{Jean2017,Koppelman2020}. Therefore, before ascribing any halo substructure to an individual accretion event, mutual associations between various substructures must be ruled out. One of the most useful tools for doing this is the study of chemical compositions of stellar atmospheres for large samples of stars. The abundance patterns of stars report back to the chemical composition of the molecular clouds they were formed from, thus constituting a fundamental indicator of their origins. Distinguishing the origin of various families of stars based on their chemical compositions is known as \textit{chemical tagging} \citep{Freeman2002}. As we will show in this section, it has been proven to be effective for dissecting stellar populations in the Milky Way's stellar halo. 

\subsection{Discovery of \textsl{Heracles}, a major building block of the Galaxy}
\label{sec_heracles}
The central few kiloparsec of the Galactic halo are extremely important when it comes to retelling the early mass assembly history of the Milky Way, and for discerning the contribution of \textit{in situ} formation to the stellar halo mass. This is because: \textit{i}) it is within the central $\sim3-4$ kpc that approximately $50\%$ of the mass of the stellar halo is contained; \textit{ii}) it is the region one would expect to find the debris from massive and/or early accreted building blocks; \textit{iii}) it is also the region one would expect to host most of the early \textit{in situ} halo star formation, including the oldest stars in the Galaxy. However, observational access to inner halo populations is quite difficult due to dust extinction and crowding by the far more numerous metal-rich stellar populations inhabiting the inner Galaxy.

\begin{figure}
\centering
\includegraphics[width=0.6\columnwidth]{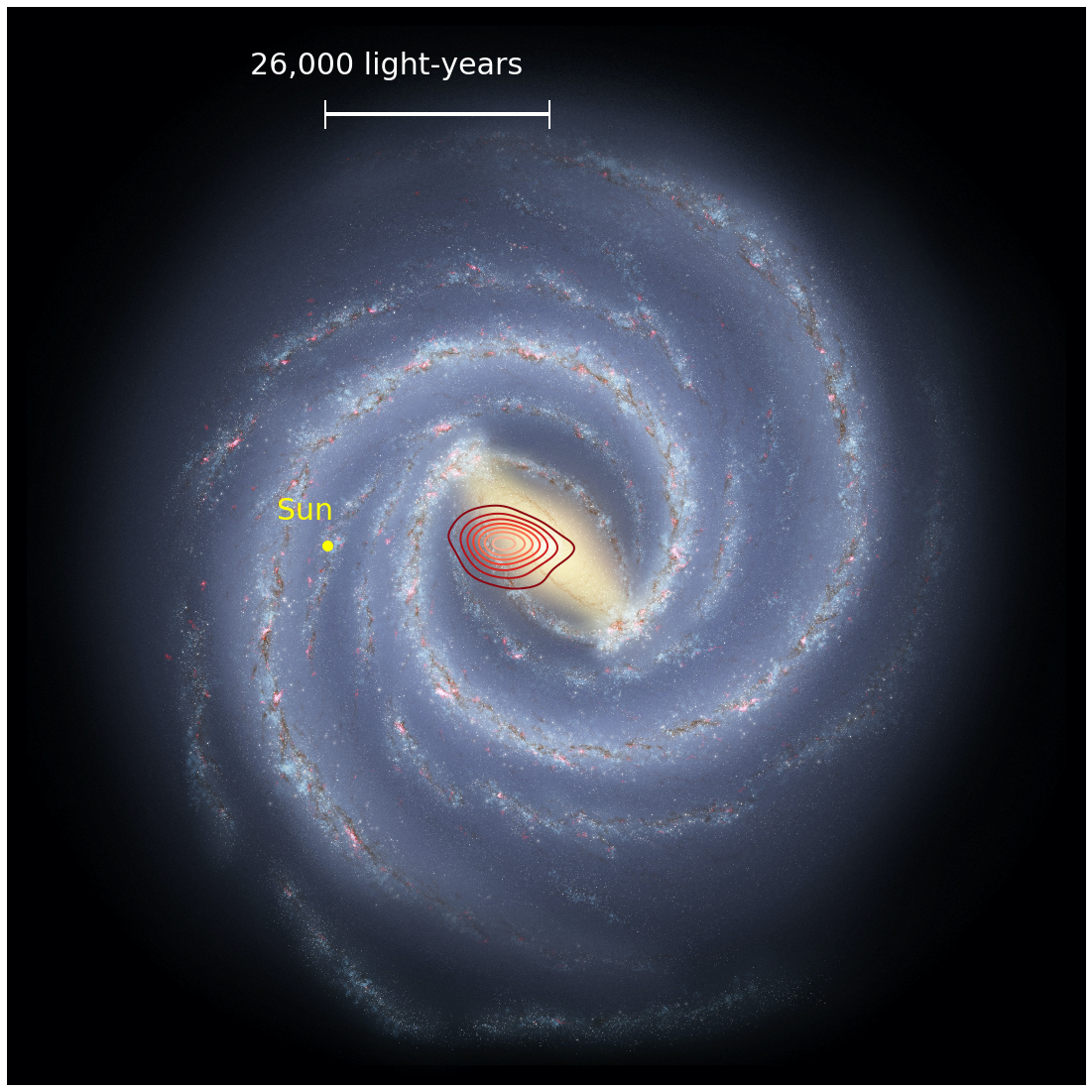}
\caption{ \footnotesize Projected spatial distribution of identified \textsl{Heracles} stars (red density contours) on the edge-on view of the MW. The stars belonging to \textsl{Heracles} were identified via a detailed analysis of its chemical-dynamical properties, and inhabit the innermost regions of the stellar halo (where most of the mass of this Galactic component is estimated to be \citealt[e.g.,][]{Horta2021b}). The chemical, orbital, spatial, and chronological properties of stellar halo populations supplied by massive stellar surveys provide an unprecedented view of the MW, and enable the unravelling of the mass assembly history of the Galaxy. \textit{Figure credit: LJMU/NASA/JPL-Caltech/Danny Horta}.}
\label{fig_heracles}
\end{figure}

Using the combination of \textsl{APOGEE} DR16 element abundance information and integrals of motion (IoM) values determined using \textsl{Gaia} DR2 astrometry 
data and calibrated distances \cite{Leung2019}, \citet{Horta2021} reported the discovery of a halo substructure buried in the heart of the Galaxy (dubbed "\textsl{Heracles}"\footnote{We note that the Heracles population was initially named Inner Galaxy Structure (IGS) upon the discovery of the stellar debris. Moreover, recent work attempting to discern the origin of the Galactic globular system, using their ages, metallicities, and dynamical properties, have also hypothesised that the inner Galaxy hosts the remnant of a major building block of the Milky Way halo \cite[e.g., \textsl{Kraken/Koala}][]{Kruijssen2020,Forbes2020}. While the association of globular clusters with stellar halo substructures is yet to be fully established, the identification of globular clusters in the heart of the Galaxy that appear to not be formed in the main progenitor of the Milky Way supports the existence of the \textsl{Heracles} debris.}, see Fig~\ref{fig_heracles}). Analysing its chemical-dynamical properties, the authors showed that this halo substructure is likely the remnant of a massive building block of the Milky Way formed at high redshift, and that appears to be chemically and dynamically detached from the more metal-rich populations with which it shares its location in the heart of the Galaxy. An examination of its [$\alpha$/Fe] and [Fe/H] abundances revealed that this population is characterised by a plateau of relatively high [$\alpha$/Fe], with the absence of a “knee” towards higher [Fe/H], indicating an early quenching of star formation activity (consistent with the scenario of an early building block of the Galaxy). Moreover, a comparison of its chemical abundance patterns with that of resembling satellite galaxies from the \textsl{EAGLE} cosmological simulations yielded an estimate of the stellar mass for the progenitor of the \textsl{Heracles} system (M$_{\star}\sim5\times10^{8}~\mathrm{M}_{\odot}$), that accounts for approximately $\sim1/3$ of the total stellar halo mass estimated to date \citep{Deason2019}. The authors also found that \textsl{Heracles} accounts for approximately $\sim1/3-1/4$ of all metal-poor (i.e., $\mathrm{[Fe/H]}<-0.8$) populations within $\sim4$ kpc from the Galactic centre, confirming predictions from different numerical simulation studies (\citealp[e.g.,][]{ElBadry2018,Fragkoudi2020,Horta2023_proto}). 

Recent observational studies have obtained results that support the presence of the \textsl{Heracles} substructure (\cite[e.g.,][]{Naidu2022,Rix2022,Arentsen2023}). Conversely, \citet{Myeong2022} used the \textsl{APOGEE-Gaia} data and suggested that \textsl{Heracles} is likely part of the \textsl{Aurora} population. However, \citet{Horta2023} quantitatively showed that the chemical abundances of \textsl{Heracles} and \textsl{Aurora} are distinct. 

The discovery of the \textsl{Heracles} population provided evidence for the existence of system in the inner Galaxy formed before the main structures inhabiting the region we understand today as the Galactic bulge were formed. It is possible that \textsl{Heracles} could have an interesting connection to the proto-Milky Way (see Section~\ref{sec_gcs}), to the presence of a population of destroyed globular cluster stars in the inner Galaxy (\citealp[][]{Schiavon2017_nrich,Kisku2021}), and the formation of the Milky Way disc.

\subsection{Chemical-dynamical characterisation of the Milky Way stellar halo}\label{sec_chemhalo}

Since the seminal work by \citet[][]{Eggen1962} and \citet{Searle1978}, many studies have aimed at characterising the stellar populations of the Milky Way, linking them to either an ``\emph{in situ}'' or accreted origin. Although detection of substructure in phase space has worked extremely well for the identification of on-going and/or recent accretion events (\citealp[e.g. Sagittarius dSph,][]{Ibata1994}; \citealp[Helmi stream,][]{Helmi1999}), the identification of accretion events early in the life of the Milky Way has proven difficult due to phase-mixing. A possible solution to this conundrum resides in the use of additional information, typically in the form of detailed chemistry and/or ages.

\begin{figure*}
\includegraphics[width=\textwidth]{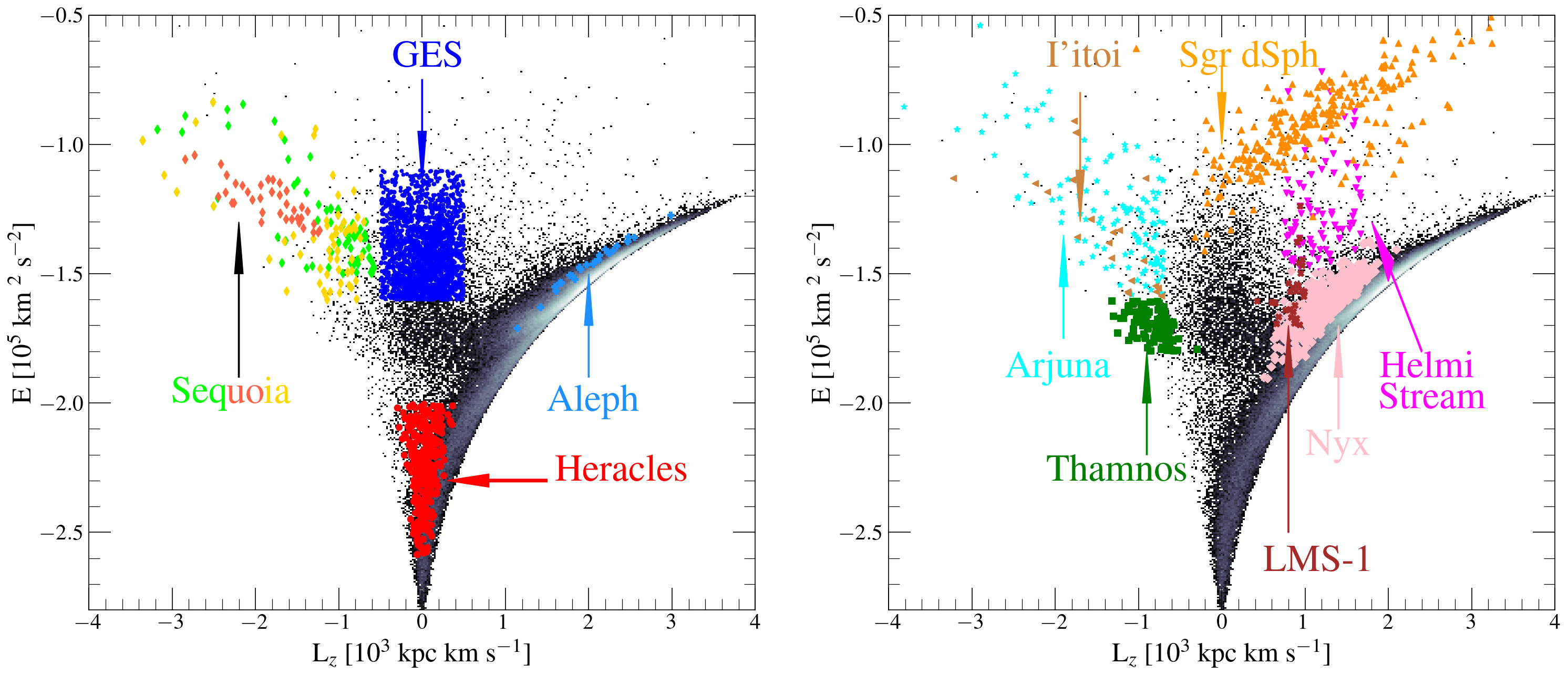}
\caption{Distribution of the identified halo substructures in the orbital energy (E) versus angular momentum w.r.t. the Galactic disc (L$_{z}$) plane. The parent sample is plotted as a 2D histogram, where white/black signifies high/low density regions. The coloured markers illustrate the different structures studied in this work, as denoted by the arrows (we do not display \textsl{Pontus}(\textsl{Icarus}) as we only identify 2(1) stars, respectively). The figure is split into two panels for clarity.}
    \label{elz}
\end{figure*}

\begin{figure*}
\includegraphics[width=\textwidth]{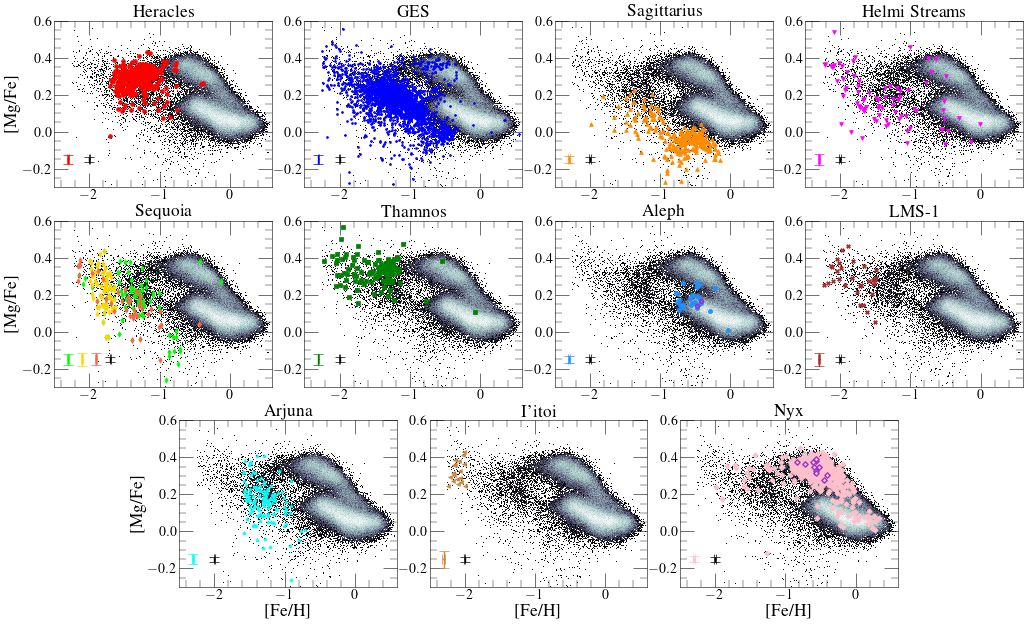}
\caption{The resulting parent sample and identified structures from Fig~\ref{elz} in the Mg-Fe plane (i.e., $\alpha$-Fe plane). The mean uncertainties in the abundance measurements for halo substructures (colour) and the parent sample (black) are shown in the bottom left corner. Colour coding and marker styles are the same as Fig~\ref{elz}. For the \textsl{Aleph} and \textsl{Nyx} substructures, we also highlight with purple edges stars from our \textsl{APOGEE} DR17 data that are also contained in the \textsl{Aleph} and \textsl{Nyx} samples from \cite[][]{Naidu2020} and \cite[][]{Necib2020} samples, respectively.}
    \label{mgfes}
\end{figure*}

\begin{figure}
\centering
\includegraphics[width=0.6\columnwidth]{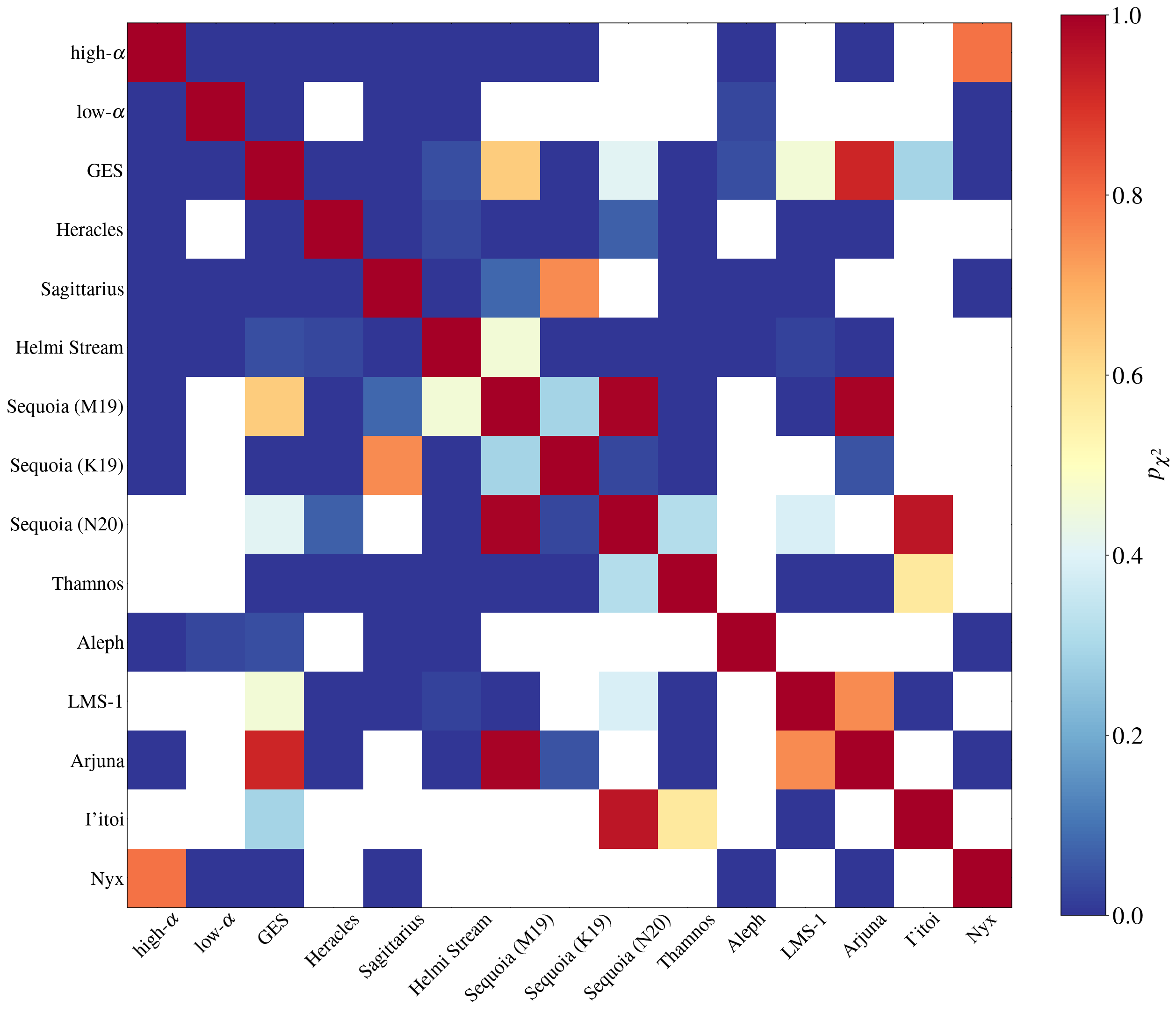}
\caption{Confusion matrix of the probability values (estimated using the $\chi^{2}$ formalism from \citealp{Horta2023}) obtained when comparing the chemical compositions of all the halo substructures with each other and with a high-/low-$\alpha$ discs. Here, each substructure is compared with its counterpart using a [Fe/H] value that is well covered by the data, where red(blue) signifies a high(low) probability of two systems being statistically similar given their chemical compositions. Comparisons with blank values are due to the two substructures being compared not having any overlap in [Fe/H].}    \label{confusion_matrix}
\end{figure}

Making use of the \textsl{APOGEE} DR17 element abundances and radial velocity information in concert with the \textsl{Gaia} DR3 astrometry, \citet{Horta2023} set out chemically characterise all the halo substructures identified dynamically with the advent of \textsl{Gaia}. The purpose of this exercise was two-fold: 1) infer the chemical evolution and star formation histories of the progenitors of defunct satellite galaxies formed at high-redshift; 2) test the reality and distinctiveness of each halo substructure. To do so, the authors first identified samples of stars belonging to the following halo substructures: \textsl{Heracles}, \textsl{Gaia-Enceladus/Sausage, Sagittarius dSph, Sequoia, Thamnos, Nyx, Arjuna, Aleph, I'itoi, LMS-1, Icarus, Cetus,} and \textsl{Pontus}. \citet{Horta2023} then set out to study qualitatively the distribution of each population in several chemical abundance planes (see Fig~\ref{mgfes} for an example), before quantitatively comparing the abundances of each system with eachother using a new statistical technique (Fig~\ref{confusion_matrix}). 

In summary, the main findings from this work can be summarised as: $i$) the chemical properties of most substructures studied match qualitatively those of dwarf Milky Way satellites, with the exception of \textsl{Nyx} and \textsl{Aleph}, whose chemical compositions resemble that of the Milky Way disc; $ii$) \textsl{Heracles} presented chemical compositions that are distinguishable from other halo substructures and from \textit{in situ} populations; $iii$) several halo substructures presented indistinguishable chemical properties from the omnipresent \textsl{Gaia-Enceladus/Sausage}, suggestive that they may be different parts of the same debris. The results from this work highlighted the importance of using chemical composition information for discerning the origin of stellar halo populations, and helped place constraints on the accretion history of the galaxy, and how we think lower-mass galaxies chemically evolve in the early Universe.

\section{Comparison to expectations from cosmological simulations} \label{sec_cossims}

Although the great strides made on the observational front are shedding light on the nature and accretion history of our Galaxy, there is still a demand to ground such observational findings on a theoretical footing. One of the best ways of performing this is via the analysis of state-of-the-art cosmological
simulations, as they provide the context to connect the Milky Way to galaxy formation theory, and now supply detailed (high-resolution) representative samples of Milky Way-mass galaxies to which survey data can be compared. This not only provides an avenue to constrain current cosmological and galaxy formation models, but also helps tackle a pivotal open question: \textit{How typical is the Milky Way?}

In this Section, we provide a summary of two studies using the \textsl{FIRE}-2 \citep{Hopkins2018} cosmological simulations aimed at grounding recent observational findings on a theoretical footing. Specifically, we aim to examine the observable properties of building blocks that constitute the Milky Way-mass galaxies.

\subsection{Unravelling the properties of proto-galaxy systems of Milky Way-mass galaxies}

Recent observational results have shown that the inner $\sim5-10$ kpc of the Galaxy's stellar halo are vital to understand the earliest stages of formation of the Galaxy, as well as the genesis of the Milky Way disc. For this reason, observational studies have set out to investigate
the properties of the oldest stellar populations in the Milky Way. These studies have reported chemical-kinematic evidence for the presence of stellar populations in the innermost regions of the Galaxy ---where one would expect the oldest stars formed \textit{in situ} to inhabit (\citealp[e.g.,][]{ElBadry2018,Fragkoudi2020})--- that are distinct from the dominant bar/disc, and are likely to constitute the entirety, or part of, the “proto-Milky Way” (\citealp[e.g.,][]{Horta2021,Belokurov2022,Rix2022}). This stellar population has been postulated to arise from the main progenitor system of the Milky Way, a major building block, or both simultaneously. The examination of this population has also shed light on the formation time and processes of the Milky Way's disc. Thus, deciphering \textit{when} and \textit{how} the proto-Milky Way formed, and examining its properties, is crucial to understand how the Galaxy became the galactic home we call the Milky Way, and for placing constraints on galaxy formation at high redshift.

\begin{figure}
\centering
\includegraphics[width=\textwidth]{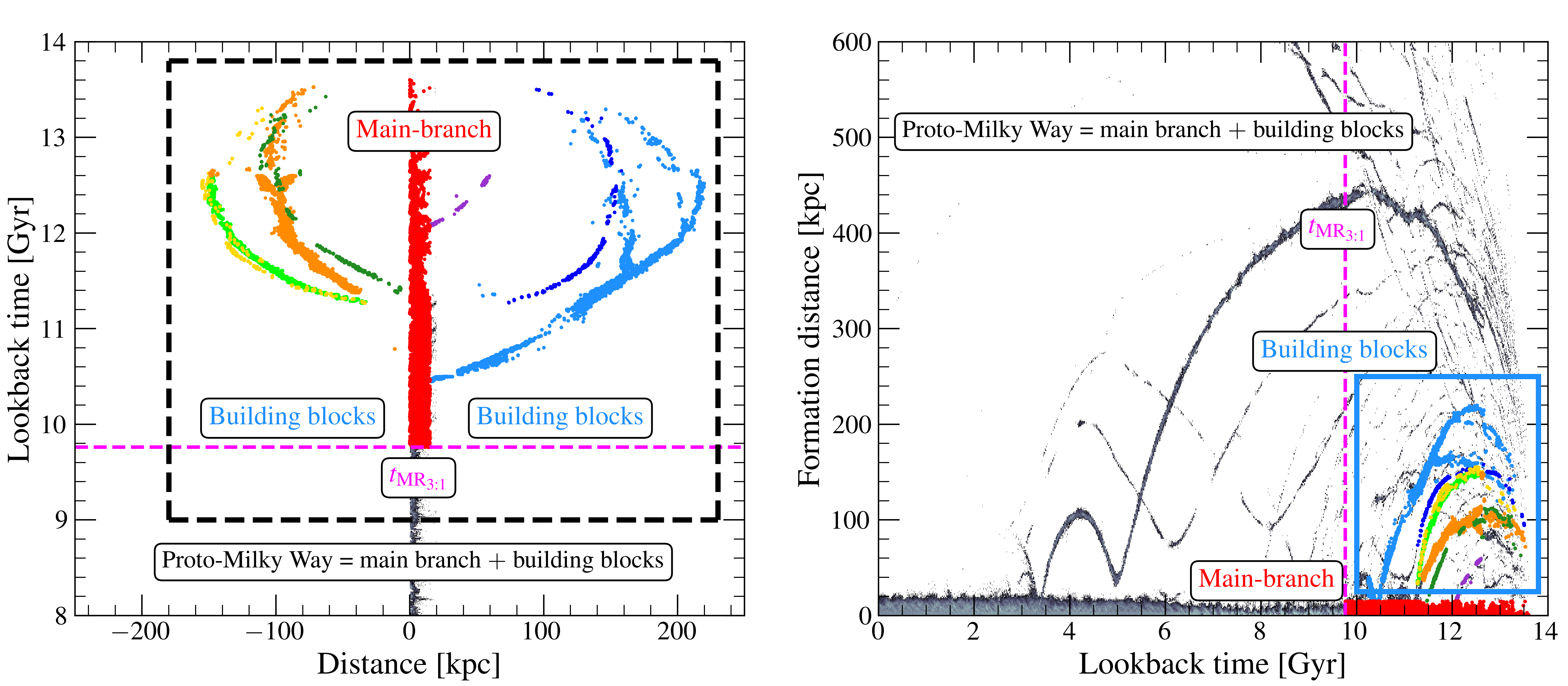}
\caption{\textit{Left}: Diagram of the merger tree of the m12b simulation in \textit{Latte} up to a lookback time of 8 Gyr. In \citet{Horta2023_proto}, a proto-Milky Way is defined as the amalgamation of the main branch halo (red) in a simulation \textit{plus} all the building blocks (other colours) that coalesce onto it before $t_{\mathrm{MR_{3:1}}}$ (dashed magenta line). \textit{Right}: Distance at which a star particle in the simulation is formed w.r.t. the centre of the main host as a function of lookback time (i.e., age) for the m12b simulation in \textit{Latte}. Highlighted in red are the star particles associated with the main branch, and as other colours the star particles associated with the resolvable building blocks (i.e., luminous subhaloes) that join with the main branch before $t_{\mathrm{MR_{3:1}}}$. By tracking each system over time, we are able to identify the star particles associated with all the proto-Milky Way fragments.}
\label{fig_proto}
\end{figure}

\begin{figure}
\centering
\includegraphics[width=0.6\columnwidth]{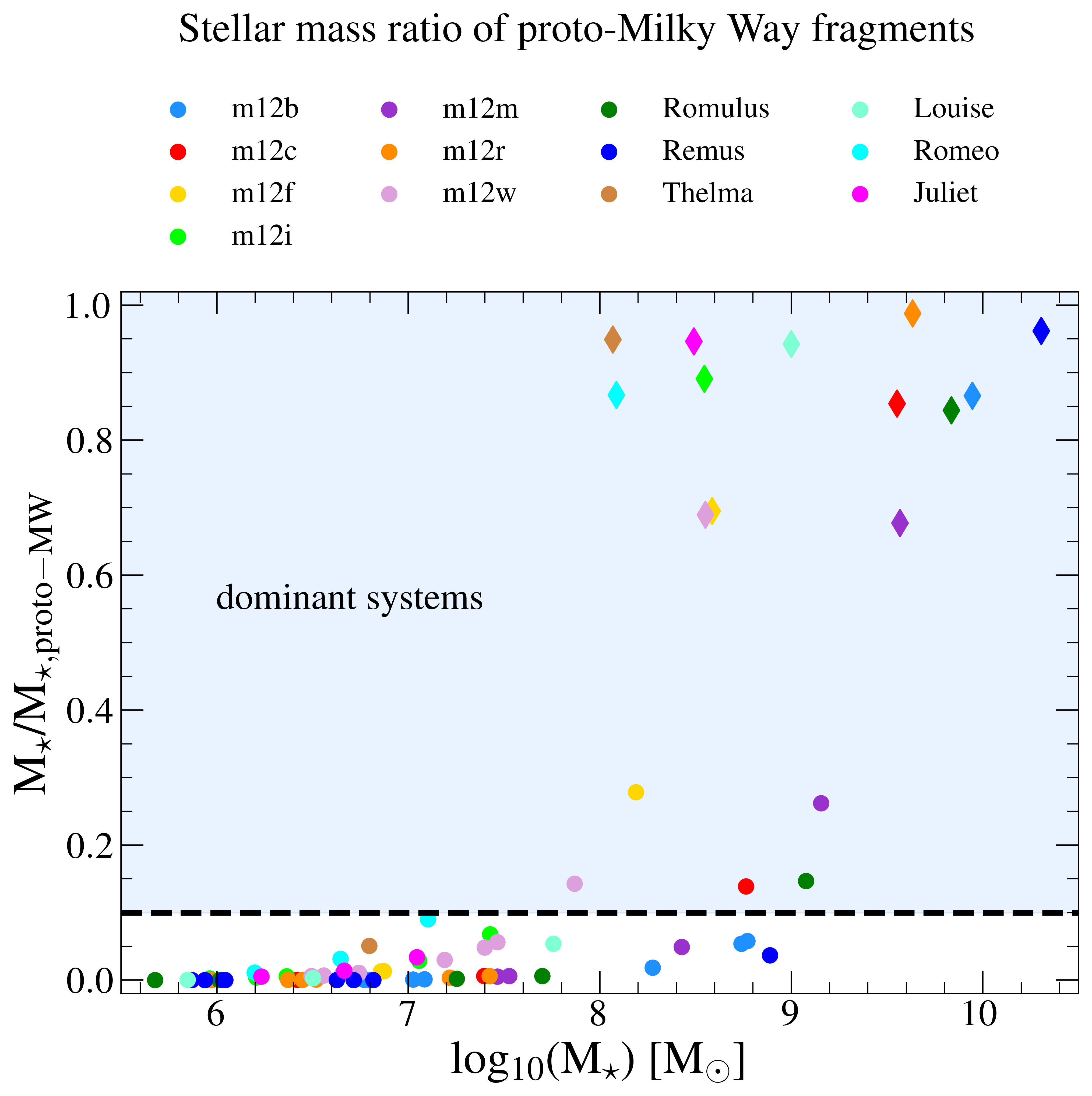}
\caption{Stellar mass ratio between each main branch (diamonds) or building block (circles) and the proto-Milky Way (i.e., main branch$+$building blocks) at $t_{\mathrm{MR_{3:1}}}$ (the time the proto-Milky Way emerges) as a function of the stellar mass of each main branch/building block. There are five clear cases (namely, m12c, m12f, m12m, m12w, and Romulus) which have a building block on the order of $\gtrsim1:5$ mass ratio with the proto-Milky Way population (i.e., blue shaded region). This indicates that $\sim40\%$ of the thirteen proto-Milky Way systems studied have a massive/dominant building block in addition to the dominant main branch halo.}
\label{proto-mass}
\end{figure}

To help tackle these questions from a theoretical front, \citet{Horta2023_proto} undertook an analysis of the properties of proto-Milky Way populations (namely, the amalgamation of the main branch halo in a simulation \textit{plus} all the building blocks before a galaxy becomes dominant in mass, see Fig~\ref{fig_proto}) using state-of-the-art high-resolution cosmological simulations (\textsl{FIRE}-2: \citep{Hopkins2018}, but see also \cite{Semenov2023}. The analysis of these populations provided a useful theoretical blueprint to guide future observational endeveours focused on deciphering the earliest stages of assembly history of the Galaxy. A particularly interesting result that emerged from this study was that $\sim40\%$ of the proto-Milky Way populations analysed were primarily formed by two dominant systems of approximately similar mass (Fig~\ref{proto-mass}). This result is particularly interesting given our discovery of \textsl{Heracles}, as it highlights that there is a possibility that the Milky Way's proto-galactic system could be primarily comprised from the debris of two major building blocks (one of them being \textsl{Heracles}; see also \citealp[e.g.,][]{Orkney2022,Garcia2023}). For more details on an in-depth study of the chemical, kinematic, and chronological properties of proto-Milky Way systems, see \cite{Horta2023_proto}.

\subsection{The observable properties of Milky Way-mass galaxy building blocks}
\label{sec_proto}

Since the papers by \citet[][]{Bullock2005} and \citet[][]{Johnston2008}, many works have sought to interpret the resulting phase-space and orbital distribution of stars in the
stellar halos of galaxies from a theoretical standpoint (\citealp[e.g.,][]{Cooper2010,Pillepich2014,Amorisco2017}), and
specifically for MW-like galaxies (\citealp[e.g.,][]{Font2011,McCarthy2012,Deason2013,Deason2015,Deason2016,Amorisco2017_mw,Dsouza2018,Monachesi2019,Evans2020,Fattahi2020,Font2020,Grand2020,Panithanpaisal2021,Cunningham2022,Dillamore2023,Khoperskov2023a,Khoperskov2023b,Khoperskov2023c,Shipp2023}). The results from such efforts have helped reveal
the diversity and complexity of substructure in the stellar halos
of galaxies like our own Milky Way, and have helped shed light on
the properties of discovered debris in the stellar halo of the
Galaxy resulting from hierarchical mass assembly. However,
many of these results were either used in: (i) the regime of
tailored/idealized N-body simulations; (ii) the context of large
cosmological simulations that do not reach the high-resolution
needed to study the intricacies of this hierarchical formation
process.

\begin{figure}
\centering
\includegraphics[width=0.8\textwidth{}{}]{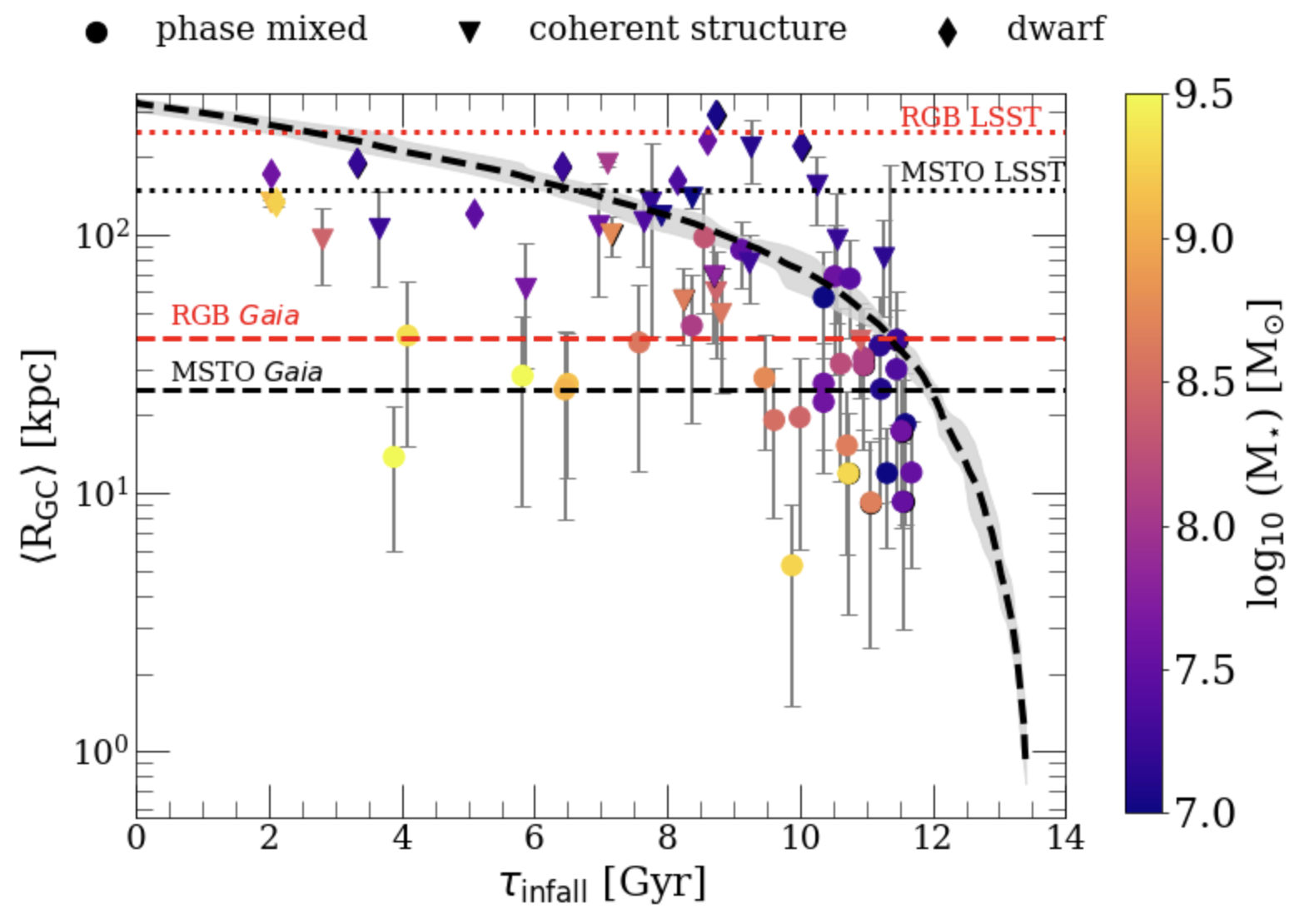}
\caption{Mean Galactocentric radius value (at $z$ = 0) for every accretion event in the seven Milky Way-mass galaxies from the \textit{Latte} suite as a function of their merger infall time (where $\tau_{\mathrm{infall}}$=0 corresponds to present day), colour coded by the stellar mass of each subhalo at the time of infall. The error bars indicate the mean of the 16$^{th}$ and 84$^{th}$ percentile ranges of the star particles. The dashed black line illustrates the mean value of the evolution of the virial radius for the seven Milky Way-mass halos used, with the shaded region demarking the 16$^{th}$ and 84$^{th}$ percentile range. Overplotted are predictions for observing detection limits for individual stars within the Milky Way galaxy, from upcoming massive surveys from \citet[][]{Sanderson2019}, assuming the faintest ($r\sim20.5-24.5$) limiting magnitudes for RGB and MSTO stars. At large $\tau_{\mathrm{infall}}$, galaxies of all stellar masses deposit the bulk of their stars close to the galactic centre of the host (MW-like) Galaxy. However, at fixed $\tau_{\mathrm{infall}}$, more massive systems tend to deposit the bulk of their stars closer to the host galaxy's centre. Furthermore, our results suggest that the smaller mass systems (M$_{\star}$ $<$ 10$^{8}$ M$_{\odot}$) that are accreted at intermediate to low $\tau_{\mathrm{infall}}$ can be contained at large galactocentric distances, of over 150 kpc, that will only be detectable with upcoming surveys like LSST.}
    \label{fig_rads_infall_mass}
\end{figure}

To test these aforementioned results and to examine the observable properties of Milky Way-mass galaxy building blocks in high-resolution cosmological simulations, \citet{Horta2023_fire} studied the observable properties of debris from building block systems in the stellar haloes of Milky Way-mass galaxies in the \textit{Latte} suite from the \textsl{FIRE}-2 cosmological simulations. To obtain a sample of every resolvable building block event across the seven Milky Way-mass haloes studied, the authors tracked the star particles belonging to building block systems from the start of the simulation until they were consumed by the resulting Milky Way-mass galaxy host. Upon obtaining this catalogue of building block debris, \citet{Horta2023_fire} set out to examine the spatial, chemical, and dynamical properties of the debris, with the aim of finding correlations between a system's intrinsic properties (e.g., stellar mass and infall time) with the observable properties studied. Their results showed that stellar mass and infall time are directly correlated with many observable properties, such as the average Galactocentric radii (Fig~\ref{fig_rads_infall_mass}), orbital energy and/or distribution in the [$\alpha$/Fe]-[Fe/H] plane. This result highlighted that properties that are observable with current stellar surveys can be used to infer the intrinsic properties of recently discovered halo substructures in the Milky Way. Moreover, one of the interesting results to emerge from this study was the relation seen between the average Galactocentric radius of the debris and the progenitors' stellar mass and infall time (Fig~\ref{fig_rads_infall_mass}).

In summary, the results from this study helped validate previous results based on lower resolution and/or $N$-body simulations. They also highlighted how observable properties (e.g., orbital energy and metallicities) can serve useful for inferring the stellar masses and infall times of galaxies that have been consumed by a Milky Way-mass host.

\section{The role of globular clusters in the formation of stellar haloes}\label{sec_gcs}

Globular clusters (GCs) are pivotal stellar populations when it comes to understanding the formation and evolution of galaxies. For the case of GCs in the Milky Way, this is because they are islands of stars, primarily formed early in the Universe, that have survived until $z=0$. This enables one to utilise GCs to study the mass assembly history of galaxies, as GCs will likely retain properties of the galaxy in which they formed, and a fraction of GCs will have been brought into the galaxy via mergers. Moreover, GCs also get destroyed during violent galactic encounters via tidal forces and stripping; they also expulse stars to the stellar halo of their host galaxies via evaporation. Thus, globular clusters donate stars to the stellar halo of their host galaxy as they evolve with time, and therefore play an important role in the formation of these galactic components. 

\subsection{Mass contribution to the total stellar halo mass budget}
The question of how the Galaxy assembled is entwined with the question of how much dissolved and/or evaporated GCs contribute to the total mass of the Galactic stellar halo. Early observational work have disclosed that in the innermost regions of the Galaxy, the contribution of dissolved and/or evaporated GCs is large ($\sim27\%$: \citealp[][]{Schiavon2017_nrich,Trincado2022}); this is not the case in the further out regions of the stellar halo ($\sim2-3\%$: \citealp[][]{Martell2010,Martell2017,Koch2019,Tang2020}). However, these estimates are solely based on number counts. 

\begin{figure}
    \centering
    \includegraphics[width=0.6\columnwidth]{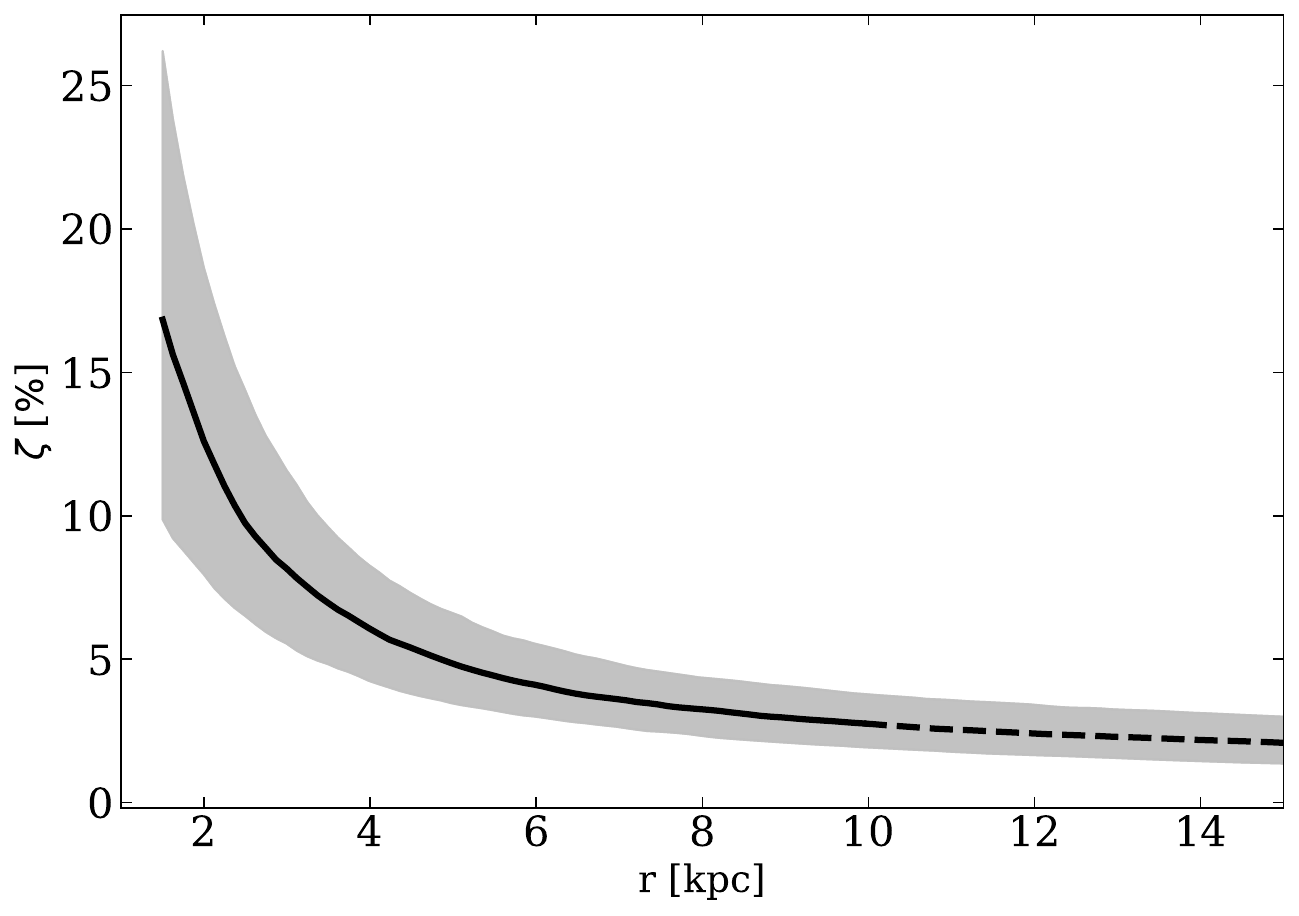}
    \caption{Mass density percentage ratio of N-rich stars and halo field stars as a function of spherical radius. The black solid line signifies the median value, while the shaded regions show the 16$^{th}$ and 84$^{th}$ percentile uncertainties. The dashed line indicates the    Galactocentric distance range where the density is not well constrained due to low numbers of N-rich stars. The mass density percentage ratio drops from $\zeta$ = 16.8$^{+10.0}_{-7.0}$$\%$ at $r = 1.5$ kpc to a value of $\zeta$ = 2.7$^{+1.0}_{-0.8}$$\%$ at $r = 10$ kpc. Under the minimal scenario assumption, one can multiply $\zeta$ by 1.5, and subtract the FP stars from the halo field, to obtain the total contribution from disrupted GC stars to the stellar halo.}
    \label{fig_nrich}
\end{figure}

To tackle this question in a more robust statistical fashion, \citet{Horta2021b} conducted a study focused on modelling the spatial distribution of stars that, based on their chemical compositions, are conjectured to be dissolved and/or evaporated GC stars in the halo field. To identify their sample, \citet{Horta2021b} applied an Extreme Deconvolution Gaussian Mixture Model (XDGMM: \citealp[][]{Holoien2017}) 
to the \textsl{APOGEE} DR16 data, and statistically identified a sample of nitrogen-rich (N-rich) stars in the Milky Way's stellar halo. The enhanced nitrogen abundances in metal-poor stars is a clear tell-tale sign that these stars likely originate from GCs, as these abundance patterns are typically seen in second generation GC populations \cite{Bastian2018}. The authors then modelled the density distribution of the N-rich star sample, and of a sample of halo field stars, testing several different density functional forms. Upon obtaining a density model that described the data well, the authors computed the fractional contribution of the N-rich star density with respect to the density of the halo field population, to estimate a fractional contribution of dissolved and/or evaporated GCs as a function of Galactocentric radius. Their results confirmed that GC dissolution processes are more prevalent in the inner $\sim2$ kpc of the Galaxy (see Fig~\ref{fig_nrich}), contributing on average $\sim28\%$ to the total stellar halo mass budget. Conversely, at larger Galactocentric distances ($r\sim10$ kpc), our findings showed that GCs contribute a much smaller average fraction of the order of $\sim4\%$. Moreover, the authors were able to estimate a total mass contribution from dissolved and/or evaporated second-generation GC stars out to $r\lesssim15$ kpc, on the order of $\sim10^{8}~\mathrm{M}_{\odot}$ (or $\sim$1/10 of the total stellar halo mass, \citealp{Deason2019}).

The question of how much destroyed GCs contribute to the total stellar halo mass budget dates back several decades \cite[e.g.,][]{Aguilar1988}. Early predictions estimated that $\gtrsim50-86\%$ of GCs have been destroyed. The results from this study help constrain these previous estimates. They also help place constraints on the mass assembly history of the Milky Way stellar halo.

\subsection{Globular clusters as tracers of the accretion history of galaxies}

In a similar vein, it is possible to use GCs tracers of the mass assembly and star formation of the Milky Way (\citealp[e.g.,][]{Forbes2010,Leaman2013,Kruijssen2019a}), as GCs retain properties of the galaxies in which they are formed. Thus, during the process of hierarchical mass assembly, the surviving GCs that are accreted onto the larger mass host will present properties that distinguish them from those formed \textit{in situ}. 

Historically, the two properties many works have used have been the ages and metallicities of GCs. However, with the advent of \textsl{Gaia} and large-scale spectroscopic surveys, there has been a push to analyse the orbits of GCs, and to use that, in combination with the age-metallicity measurements, to infer the origin of the Galactic GC system.

\begin{figure}
    \centering
    \includegraphics[width=1\columnwidth]{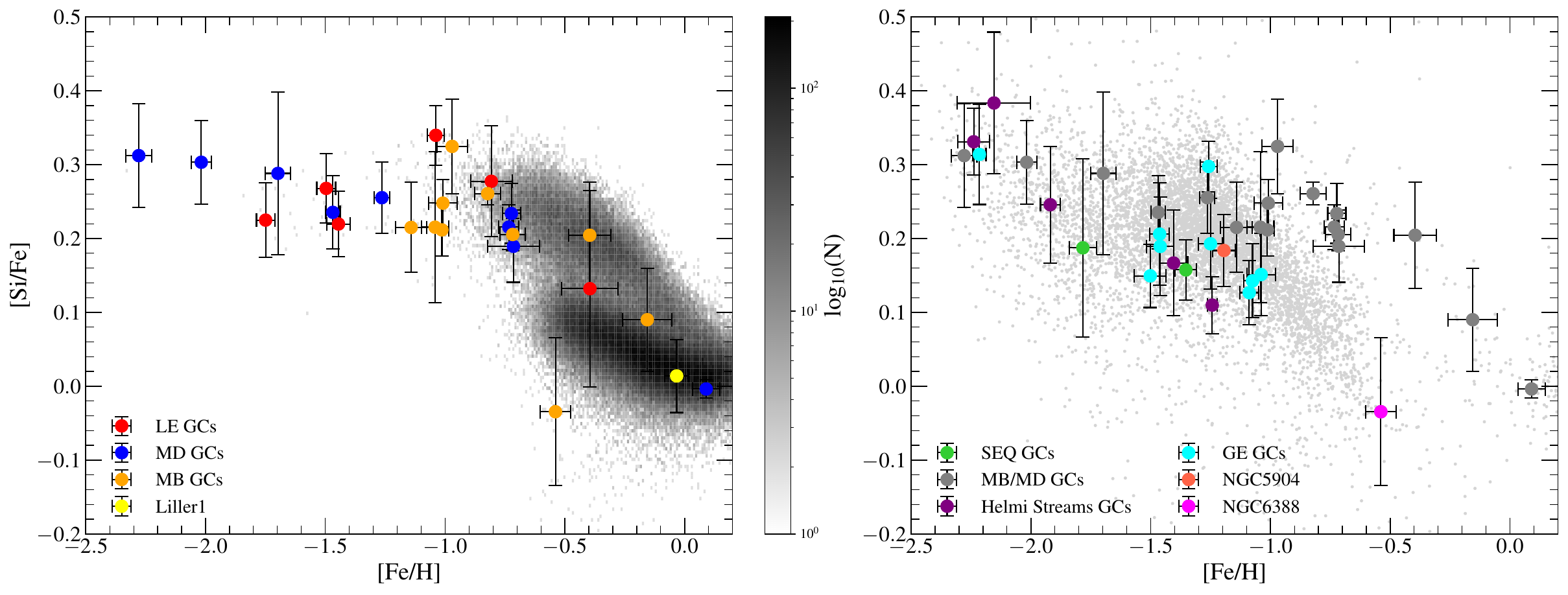}
    \caption{\textit{Left}: Mean [Si/Fe] vs [Fe/H] chemical-abundances for the Low Energy (LE, red), Main Bulge (MB, orange) and Main Disc (MD, blue) GC subgroups, illustrated alongside Liller\,1 (yellow), with the
		    1~$\sigma$ spread represented in black error bars. As grey, we show the Galactic disc and bulge field populations defined kinematically according to \citet{Massari2019}. From
		    these abundance plots, by accounting for the $1\sigma$ spread uncertainties, we find that the more [Fe/H] rich LE GCs, namely NGC\,6121, NGC\,6441 and Pal\,6 can be categorized to be from \textit{in situ} origin. The other three LE GCs still occupy the same locus as the MD/MB subgroups, however due to their low [Fe/H] abundances and dynamical properties, it is possible that these more metal-poor GCs could be from an accreted origin. Furthermore, we find that Liller\,1 occupies the same locus as the \textit{in situ} GCs, which coupled with its high [Fe/H] value can be classified as a MB GC. \textit{Right}: The same as left but for the GE (cyan), Seq (green), H99 (purple) and MD/MB (grey) GC subgroups, illustrated alongside NGC\,5904 (red) and NGC\,6388 (magenta), with the $1\sigma$ spread represented in black error bars. As light-grey points we show the halo field population. The GE, Seq and H99 accreted dwarf spheroidal subgroups occupy the same locus, displaying lower mean [Si/Fe] values to the GCs from the MD and MB populations at the same metallicity range $-1.5 < \mathrm{[Fe/H]} < -1$. According to galaxy chemical-evolution models, this suggests that either: both accreted dwarf spheroidals must have had a similar chemical-evolution history and therefore have been of similar mass, or that some, possibly all, originate from the same accretion event.  Below $\mathrm{[Fe/H]} < -1.5$, the \textit{in situ} and accreted groups are indistinguishable in the Si-Fe plane (see also \cite{Pagnini2023}). NGC\,288 displays higher [Si/Fe] values than the rest of the GE subgroup GCs ($\sim 0.15$ dex greater) of similar metallicity, however displays a clear accreted-like orbit. NGC\,5904 clearly occupies the same locus as the accreted population of GCs. However, due to the uncertainties in the measurements, it is impossible to suggest to which accreted subgroup NGC\,5904 belongs to. Along the same lines, NGC\,6388 occupies the same locus as the [Fe/H]-rich halo field population, which coupled with its retrograde orbit hints that this GC belongs to an accreted subgroup.}
    \label{fig_gcs}
\end{figure}

In a study attempting to place further constraints on both the origin of the Galactic GC system and the mass assembly history of the Galaxy, \citet{Horta2020} used the \textsl{APOGEE} DR16 data coupled with \textsl{Gaia} DR2 to obtain a homogeneous GC star catalogue. The aim of this study was to obtain a sample of stars belonging to many Galactic GCs for which precise chemical abundance information was available, to then study the chemical compositions of GCs postulated to be formed \textit{in situ} from those that are believed to be accreted. To do so, \citet{Horta2020} examined the [$\alpha$/Fe]-[Fe/H] plane, using Si as their $\alpha$ tracer element (other $\alpha$ elements like Mg can vary in GCs due to multiple populations). Their results showed that the majority of accreted GCs displayed lower [$\alpha$/Fe] abundances, at fixed [Fe/H], when compared to those postulated to be formed in situ GCs (see Fig~\ref{fig_gcs}), following that of their field counterparts. These results also helped categorise previously unclassified GCs (e.g., Liller 1). 

In summary, this work showed that, in addition to ages and orbits, the detailed element abundance patterns of GCs can be used to infer the origin of the Galactic GC system. In turn, these findings also help place constraints on the origin of specific GCs, and on the mass assembly history of the Milky Way.

\clearpage

\section*{Concluding remarks}

All together, the results summarised in this article highlight the importance of disentangling Galactic stellar halo populations to unveil the past and mass assembly history of the Galaxy. While these results brought about fresh insights on the assembly history of the Milky Way, there are key questions that still remain unanswered. One vital open question is "\textit{How many major accretion events has the Milky Way undergone?"} In the last decade, many studies have reported the discovery of halo substructures postulated to be the debris from individual disrupted satellite galaxies (\citealp[e.g.,][]{Belokurov2018,Helmi2018,Myeong2019,Koppelman_thamnos,Naidu2020,Yuan2020,Refiorentin2021,Malhan2022}), including the \textsl{Heracles} population presented in Section~\ref{sec_heracles}. However, many of these findings are based solely on orbital information from \textsl{Gaia}, which can lead to erroneous assignments to new accretion events of substructures that in fact are associated to already known progenitors \citep[e.g.,][]{Jean2017,Koppelman2020}. In Section~\ref{sec_chemhalo} we showed the importance of using chemical abundance information to ascertain the reality of halo substructures, and via a chemical abundance study showed that many halo substructures resemble the omnipresent \textsl{Gaia-Enceladus/Sausage} accretion and/or populations formed \textit{in situ}. Our findings have helped elucidate the reality of these halo populations, and in summary suggest that the Milky Way likely has undergone three major accretions (namely, \textsl{Heracles}, \textsl{Gaia-Enceladus/Sausage}, \textsl{Sagittarius} dSph) and a number of lower-mass ones (\textsl{Thamnos, Helmi Stream}). However, there are still many steps that need to be taken before establishing a full picture of the accretion history of the Galaxy.

More specifically concerning the nature of \textsl{Heracles}, another important question that needs answering is "\textit{Is Heracles fully contained within the inner $\sim4$ kpc from the Galactic Centre as reported, or do the debris span wider spatial regimes?}" Theoretical expectations suggest that whilst the bulk of the debris from an early and massive cannibalised satellite galaxy (as \textsl{Heracles} is likely to be) may be predominantly contained within small Galactocentric radii, there should be some fraction of stars that occupy larger distances (see Fig~\ref{fig_rads_infall_mass}). Assessing the true spatial extent of \textsl{Heracles} debris is a key question that shall hopefully be answered with the advent of SDSS-V. Along those lines, other obvious questions that still remain unsolved are "\textit{How many accreted debris reside in these innermost $\sim4$ kpc of the Galactic stellar halo?} and "\textit{Is Heracles (and other early building block systems) distinguishable in chemical-dynamical space from the oldest in situ halo populations in the Milky Way's stellar halo?}" The answer to these two questions will shed light on the earliest stages of formation of the Galaxy. So far, the latest observational data support the notion that a major building block of the Galaxy inhabits these innermost regions (i.e., \textsl{Heracles}). However, the oldest \textit{in situ} halo must also reside in this region. Moreover, results from cosmological simulation studies (see Section~\ref{sec_cossims}) suggest that building blocks like \textsl{Heracles} occur in Milky Way-mass galaxies, and are not uncommon (see Fig~\ref{fig_proto} and Fig~\ref{proto-mass}). In a similar vein, studies on the properties of Galactic GCs predict that a building block like \textsl{Herackes} should have occured in the Milky Way (i.e., ``\textsl{Kraken}''/``\textsl{Koala}'' \citealp[]{Kruijssen2020,Forbes2020}). Whilst \textsl{Heracles} overlaps in the E-L$_{z}$ diagram with this predicted system, a definitive connection between them is yet to be established.

Interestingly, there may be a connection between the fact that there is evidence for an accreted debris in the innermost Milky Way halo, and that the incidence of disrupted/evaporated globular cluster stars (i.e., N-rich stars) increases by a factor of $\sim$5 in this region when compared to the outer halo. \citet{Kisku2021} showed that many of these N-rich stars present chemical-dynamical features that are consistent with \textsl{Heracles}. Thus, it is possible that \textsl{Heracles} either brought in GCs which were destroyed during the merger, or that the \textsl{Heracles}-Milky Way interaction caused \textit{in situ} GCs in this region to be destroyed at a higher rate than in the outer halo. Moreover, theoretical expectations suggest that Milky Way-mass haloes that resemble the Galaxy chemically (i.e., present an $\alpha$-bimodality) assemble their mass early \citep{Mackereth2018} and typically have a higher number of disrupted/evaporated globular cluster stars in their inner ``bulge'' region \cite{Hughes2020}, in line with the results presented here. All in all, it could be that these two findings may be related. In addition, these results have ramifications for our understanding of the origin of the Galactic GC system. Thus, chemically/chronologically/dynamically characterising Galactic GCs, as done in Section~\ref{sec_gcs} and in previous works (\citealp[e.g.,][]{Forbes2020,Kruijssen2020,Callingham2022,Monty2023}), should also help elucidate the assembly history of the Galaxy. However, deciphering which property (orbits, chemical compositions, or ages) is the most reliable tracer seems to be a necessary first step (Monty et al., in prep).

\clearpage

\backmatter

\bmhead{Acknowledgments}
DH would like to thank the referee for a constructive report. He would also like to thank his family, Sue, Alex, and Debra, for all the support during the years in which all this work was conducted. The authors would like to thank Ted Mackereth, Nate Bastian, Sebastian Kamann, and the \textsl{APOGEE} and \textsl{FIRE} teams for useful discussions during the conduction of the research leading up to this paper.

\begin{appendices}




\end{appendices}

\bibliography{biblio}


\end{document}